\documentclass[10pt,aps,twocolumn,prd,
superscriptaddress,noshowpacs,nofootinbib,noshowkeys,floatfix
]{revtex4-1}
\usepackage[dvips]{graphics,graphicx}
\usepackage[colorlinks=true,linktocpage=true,linkcolor=blue,citecolor=blue]{hyperref}
\usepackage[usenames,dvipsnames]{color}
\usepackage{amsmath, amssymb}
\usepackage{multirow}
\usepackage{longtable}
\usepackage{color}
\usepackage[normalem]{ulem}  

\renewcommand\sout{\bgroup \color{blue} \ULdepth=-.5ex \ULset}

\newcommand{\taueq}{\tau_R}
\newcommand{\pres}{{\cal P}}
\newcommand{\ped}{{\cal E}}
\newcommand{\pedeq}{\ped_0}
\newcommand{\preseq}{{\pres_0}}
\newcommand{\deltaf}{\delta\! f}
\newcommand{\deltaB}{\delta\! B}
\newcommand{\eq}{\text{0}}
\newcommand{\VP}{\vphantom{\frac{}{}}}

\def\dM{\kappa}
\def\hT{{\hat T}}

\begin{document}
 
\preprint{}

\title{Quasiparticle second-order viscous hydrodynamics from kinetic theory}

\author{Leonardo Tinti}
\affiliation{Department of Physics, The Ohio State University, Columbus, OH 43210-1117, USA}
\author{Amaresh Jaiswal}
\affiliation{GSI, Helmholtzzentrum f\"ur Schwerionenforschung, Planckstrasse 1, D-64291 Darmstadt, Germany}
\author{Radoslaw Ryblewski}
\affiliation{Institute of Nuclear Physics, Polish Academy of Sciences, PL-31342 Krak\'ow, Poland}
 
\date{\today}

\begin{abstract}

We present the derivation of second-order relativistic viscous hydrodynamics from 
an effective Boltzmann equation for a system consisting of 
quasiparticles of a single species. We consider temperature-dependent masses of the quasiparticles and devise a thermodynamically-consistent 
framework to formulate second-order evolution equations for shear 
and bulk viscous pressure corrections. The main advantage of this formulation is 
that one can consistently implement realistic equation of state of the 
medium within the framework of kinetic theory. Specializing to the 
case of one-dimensional purely-longitudinal boost-invariant expansion, we 
study the effect of this new formulation on viscous hydrodynamic 
evolution of strongly-interacting matter formed in relativistic 
heavy-ion collisions.

\end{abstract}



\maketitle 

\section{Introduction} 
%
Relativistic fluid dynamics has been widely applied to study 
space-time evolution of the strongly-interacting, hot and dense 
matter created in ultra-relativistic heavy-ion collisions at the 
Relativistic Heavy-Ion Collider (RHIC) in BNL and the Large Hadron 
Collider (LHC) in CERN; for reviews see~\cite{Huovinen:2006jp, 
Romatschke:2009im, Gale:2013da, Jaiswal:2016hex, 
Jeon:2016uym}.~Multiple successes of the pioneering studies 
employing ideal fluid dynamics to describe experimental data at RHIC led to establishing the paradigm of perfect-fluidity of the created new phase of matter called the quark-gluon plasma (QGP) \cite{Shuryak:2003xe} and raised questions about its fast thermalization \cite{Heinz:2001xi}; 
see also \S1.3 in Ref.~\cite {Ryblewski:2013jsa}.~The universal  existence of the viscous effects in nature \cite{Danielewicz:1984ww,Policastro:2001yc,Kovtun:2004de} as well as the increasing 
precision of flow measurements at the LHC suggested necessity of 
inclusion of the dissipative effects in the fluid dynamic modelling and resulted in a rapid development of the formulation of relativistic viscous hydrodynamics \cite{Muronga:2003ta, 
Baier:2006um, York:2008rr, El:2009vj, Denicol:2010xn, 
Denicol:2012cn, Denicol:2012es, Jaiswal:2012qm, Jaiswal:2013npa, 
Jaiswal:2013fc, Jaiswal:2013vta, Bhalerao:2013aha, Bhalerao:2013pza, Denicol:2014vaa, Chattopadhyay:2014lya, Kikuchi:2015swa, 
Tsumura:2015fxa}. Much of the research in the field of heavy-ion 
collisions has been devoted to the extraction of the thermodynamic 
and transport properties of the QGP medium encoded mainly in its equation 
of state, shear viscosity $\eta$, and bulk viscosity $\zeta$ \cite{Romatschke:2007mq, Bozek:2009dw, Song:2010mg, Alver:2010dn,Schenke:2011bn,Gale:2012rq,Noronha-Hostler:2013gga,Gardim:2014tya,Ryu:2015vwa,Niemi:2015qia}.

The successes of viscous fluid dynamics in explaining a wide range 
of collective phenomena observed in high-energy heavy-ion collisions 
have been initially attributed to the proximity of the created QGP 
state to the local thermodynamic equilibrium.~This assumption plays 
a key role in the formulation of the theory of relativistic 
hydrodynamics as it is usually constructed as an order-by-order 
expansion around equilibrium state in powers of thermodynamic 
gradients, where ideal hydrodynamics is of zeroth order.~The 
first-order relativistic Navier-Stokes theory \cite{Eckart:1940zz, 
Landau:1959} involves parabolic differential equations and therefore 
suffers from acausality and instability. The second-order 
Israel-Stewart theory \cite{Israel:1976tn, Stewart:1977, 
Israel:1979wp} leads to hyperbolic equations which restores 
causality but may not guarantee stability \cite{Lindblom:1995gp, 
Geroch:1995bx}.

The procedure outlined above indirectly sets the formal 
applicability limits of the resulting theory, making it questionable 
when the dissipative corrections are large \cite 
{Strickland:2014pga}.~Interestingly, recent findings within the 
gauge/gravity duality framework \cite{Heller:2011ju}, as well as, in 
the effective kinetic theory approach \cite{Kurkela:2015qoa} show 
that, in practice, the relativistic viscous hydrodynamics behavior 
in heavy-ion collisions sets in quite early, even though the system 
is far from equilibrium -- the phenomenon commonly known as the 
hydrodynamization of the system.~These observations suggest that the 
applicability of the viscous hydrodynamics may be broader than 
previously expected.

In order to describe the collective behaviour of the QGP within 
fluid dynamics, one needs to incorporate its properties through the 
transport coefficients and equation of state.~In principle, the two 
should be treated as an external input in the  hydrodynamic 
equations, and extracted from the experimental data.~As, in general, 
it is a highly non-trivial task, one usually follows different 
methodology.~Since the full information on the properties of nuclear 
matter produced in heavy-ion collisions should follow from the 
fundamental theory of strong interactions, namely Quantum 
Chromodynamics (QCD), one may try to incorporate into hydrodynamics 
the results of ab-initio calculations of thermodynamic and transport 
quantities performed in the framework of lattice QCD (lQCD) \cite 
{Fodor:2009ax, Bazavov:2009zn,Borsanyi:2010cj, Bazavov:2014pvz, 
Borsanyi:2013bia}.~However, it turns out that in contrast to the 
classical Navier-Stokes theory the form of the relativistic viscous 
hydrodynamics equations is not universal and strongly depends on the 
underlying microscopic theory used to derive them \cite 
{Muronga:2003ta, Baier:2006um, York:2008rr, El:2009vj, 
Denicol:2010xn, Denicol:2012cn, Denicol:2012es, Jaiswal:2012qm, 
Jaiswal:2013npa, Jaiswal:2013fc, Jaiswal:2013vta, Bhalerao:2013aha, 
Bhalerao:2013pza, Denicol:2014vaa, Chattopadhyay:2014lya, 
Kikuchi:2015swa, Tsumura:2015fxa, Florkowski:2013lza, 
Florkowski:2013lya, Florkowski:2014sfa, Jaiswal:2014isa, 
Denicol:2014mca, Florkowski:2015lra}.~Moreover, in the 
phenomenologically interesting regime the lQCD calculations of 
transport properties are still plagued by large uncertainties \cite 
{Meyer:2007ic, Meyer:2007dy}.

In view of the above problems, one typically resorts to a simple 
microscopic theory, such as the kinetic theory, to derive the 
hydrodynamic evolution equations. Subsequently, in order to 
 incorporate realistic properties of the strongly-interacting 
matter in the hydrodynamic evolution, equation of state and 
transport coefficients obtained from lQCD are implemented. 
However, the latter procedure inadvertently fixes the parameters of the 
microscopic theory, introducing effective interactions which were 
not taken into account in the derivation of the hydrodynamic 
evolution equations. These inconsistencies eventually lead to the 
violation of thermodynamic relations in such a system.

In the specific case of relativistic kinetic theory for a single 
particle species, the equation of state is fixed by the particle 
mass. Using this simple formalism it is not possible to fit exactly 
the temperature scaling of energy density and pressure given by lQCD 
as one may only select the most convenient effective mass. In order 
to improve the fit, one may consider temperature (medium) dependent 
particle masses, which, in a thermodynamically consistent framework, lead to a non-ideal equation of state \cite{Gorenstein:1995vm,Romatschke:2011qp,Sasaki:2008fg,Bluhm:2010qf}. This is achieved by introducing an 
extra contribution to the energy-momentum tensor which can be 
physically interpreted as a mean field. While this procedure is 
equivalent to the introduction of the notion of interacting 
quasiparticles in the microscopic theory, only in specific limits 
they can  be considered actual quasiparticle excitation of the 
fundamental theory~\cite{Romatschke:2011qp}.

Once an effective quasiparticle kinetic theory is constructed which 
can reproduce any thermodynamically consistent equation of state, 
hydrodynamic evolution equations should be derived by coarse 
graining this theory. To the best of our knowledge, this has not yet 
been done. In this paper, we derive evolution equations for 
second-order viscous hydrodynamics for a system of single species of 
quasiparticles from an effective Boltzmann equation in the 
relaxation time approximation. Specializing to the case of 
transversally homogeneous and boost-invariant longitudinal 
expansion, we study the effect of this new formulation on the 
evolution of strongly interacting matter formed in high energy 
heavy-ion collisions.

\section{Quasiparticle thermodynamics}

Let us consider a system of ideal (non-interacting) uncharged 
particles of a single species. Within kinetic theory, the equation 
of state of such a system depends parametrically only on the mass of 
the particle \cite{Landau:1975}. In order to describe an arbitrary 
equation of state, one can consider temperature-dependent masses of 
the particles, $m(T)$ \cite{Goloviznin:1992ws}. Such an idea may be 
physically sound, for instance, when considering high-temperature 
QCD. In this case resummed perturbative calculations suggest that 
the system is made of partons with thermal masses $m(T)= g_s T$, where 
$g_s$ is the strong coupling \cite{Braaten:1989mz}. However, these 
particles do not correspond to any real excitations of the 
underlying fundamental theory, especially close to the crossover 
region. While introducing a temperature-dependent mass leads to some 
degree of freedom in choosing the equation of state of the system, 
unfortunately, it violates basic thermodynamic identities \cite
{Gorenstein:1995vm}.

A possible way to restore thermodynamic consistency at global 
equilibrium is to introduce additional effective mean field through 
a bag function $B_0(T)$ to account for the fundamental interactions 
giving rise to the in-medium masses \cite{Gorenstein:1995vm}. The 
latter may be included in the Lorentz covariant way by modifying the 
definition of the equilibrium energy-momentum tensor \cite
{Jeon:1994if, Jeon:1995zm, Chakraborty:2010fr, Romatschke:2011qp, 
Albright:2015fpa}
\begin{equation}\label{Tmunu_eq}
T_\eq^{\mu\nu} = \int dP \, p^\mu p^\nu \, f_\eq  \, + B_0(T) \, g^{\mu\nu},
\end{equation}
where $g^{\mu\nu}=\rm{diag}(1,-1,-1,-1)$ is the metric tensor, 
$f_\eq$ is the equilibrium distribution function, $p^\mu$ is the 
quasiparticle four-momentum and $B_0(T)$, called the bag 
pressure/energy, is a function which can be determined by requiring 
thermodynamic consistency. In the above equation, $dP$ is the 
Lorentz covariant momentum integration measure defined as
\begin{equation}\label{int_mes}
\int dP  =  \int \frac{d^4 p}{(2\pi)^4} \, 2 \, \Theta(p\cdot t) \, (2\pi) \, \delta(p^2 - m^2),
\end{equation}
where $\Theta$ is the Heaviside step function, $t^\mu$ is an 
arbitrary time-like four-vector, $p\cdot t \equiv p^\mu g_{\mu\nu} 
t^\nu$ and $p^2 \equiv p \cdot p$.


The equilibrium energy density and pressure can be defined as
\begin{equation}\label{eng_prs}
\pedeq = u\cdot T_\eq \cdot u, \quad \preseq = -\frac{1}{3}\Delta : T_\eq ,
\end{equation}
where $\Delta^{\mu\nu}=g^{\mu\nu}-u^\mu u^\nu$ with $\Delta^{\mu\nu} 
u_\nu =0 $ and $u^\mu$, which will be defined in Sec.~\ref{sec:non}, 
is the fluid four-velocity satisfying $u^2=1$. We also introduced 
the following notation $A:B\equiv A^{\mu\nu}B_{\mu\nu}$ for the 
Frobenius product. The thermodynamic relation
\begin{equation}\label{th_rel}
\frac{d\preseq}{dT} = \frac{\pedeq + \preseq}{T},
\end{equation}
is guaranteed to be satisfied if \cite{Gorenstein:1995vm}
\begin{equation}\label{equilibrium_bag}
dB_0 + m \,d m \int dP \; f_\eq = 0.
\end{equation}
The latter relation can be seen explicitly for the case of 
Maxwell-Boltzmann distribution~\cite{deGroot:1980}, 
\begin{equation}\label{fMB}
f_\eq=g\exp[-\beta(u \cdot p)],
\end{equation}
where $g$ is the degeneracy factor and $\beta\equiv1/T$. In this case the  equilibrium energy density and pressure is given by
\begin{align}\label{eng_den_B}
\pedeq &= \frac{g T^4 z^2}{2\pi^2} \left[\VP3 K_2(z) + z K_1(z) \right] + B_0 \\\label{prs_B}
\preseq &= \frac{gT^4 z^2}{2\pi^2}   K_2(z) - B_0,
\end{align}
where $K_n(z)$ are the modified Bessel functions of second kind of 
order $n$ and $z \equiv m/T$. The temperature derivative of pressure is 
given by
\begin{align}
\frac{d\preseq}{dT} &= \frac{g T^3 z^2}{2\pi^2} \left[ 4 K_2(z) + z K_1(z) -\frac{dm}{dT} \, K_1(z) \right] - \frac{dB_0}{dT} \nonumber\\
&= \frac{\pedeq +\preseq}{T} - \left(\vphantom{} \frac{dB_0}{dT} + m \, \frac{dm}{dT}\int dP f_\eq \right),
\end{align}
which leads to Eq.~(\ref{equilibrium_bag}) in order to satisfy the 
thermodynamic relation given in Eq.~(\ref{th_rel}).

\section{Non-equilibrium and Boltzmann equation}
\label{sec:non}

\subsection{Off-equilibrium mean field and the energy-momentum conservation}
\label{ssec1:non}

For the general non-equilibrium case, we propose the energy-momentum 
tensor of the form
\begin{equation}\label{Tmunu_noneq}
 T^{\mu\nu} = \int dP \, p^\mu p^\nu \, f  \; + B^{\mu\nu},
\end{equation}
where in equilibrium we require $B^{\mu\nu}|_{\rm eq} = B_0 g^{\mu\nu}$. With $f\to f_\eq$, 
the above equation reduces to Eq.~(\ref{Tmunu_eq}) in equilibrium. 
We now split $B^{\mu\nu}$ into equilibrium and non-equilibrium parts,
\begin{equation}\label{Bneq}
B^{\mu\nu} = B_0 \, g^{\mu\nu} + \deltaB^{\mu\nu}, 
\end{equation}
where $\deltaB^{\mu\nu}$ has to be fixed by requiring 
energy and momentum conservation. The dissipative quantities are defined 
as
\begin{align}
 \Pi &\equiv -\frac{1}{3}\Delta : \left( T  - T_\eq \right), \label{bulk_def}\\
 \pi^{\mu\nu} &\equiv \Delta^{\mu\nu}_{\alpha\beta} \left( T^{\alpha\beta} 
 - T^{\alpha\beta}_\eq \right) = \Delta^{\mu\nu}_{\alpha\beta} \, T^{\alpha\beta} \label{shear_def},
\end{align}
where $\Pi$ is the bulk, $\pi^{\mu\nu}$ is the shear pressure 
correction, and $\Delta^{\mu\nu}_{\alpha\beta}=\frac{1}{2} 
(\Delta^{\mu}_{\alpha}\Delta^{\nu}_{\beta} + 
\Delta^{\mu}_{\beta}\Delta^{\nu}_{\alpha} - \frac{2}{3} 
\Delta^{\mu\nu}\Delta_{\alpha\beta})$ is the symmetric traceless 
projector orthogonal to $u^\mu$.

The energy-momentum conservation equation requires that the four 
divergence of energy-momentum tensor should vanish, i.e., 
$\partial_\mu T^{\mu\nu}=0$.~The four-divergence of $T^{\mu\nu}$ in  
Eq.~(\ref{Tmunu_noneq}) reads
\begin{align}
 \partial_\mu T^{\mu\nu} = &~ \partial_\mu B^{\mu\nu} + \!\int\!\! \frac{d^4 p}{(2\pi)^3}\, 
 2\Theta(p\cdot t)(\VP p\cdot \partial)\delta(p^2 - m^2) \,  p^\nu f \nonumber\\
 & + \int dP \, p^\nu (p\cdot \partial) f.
\end{align}
Rewriting the Dirac delta function in the above equation,
\begin{equation}
 (\VP p\cdot \partial) \delta(p^2 -m^2) = m\,(\partial_\mu m) \left[ -\partial^\mu_{(p)}\delta(p^2-m^2) \right],
\end{equation}
where $\partial^\mu_{(p)}$ is the gradient with respect to the 
quasiparticle momenta, the energy-momentum conservation leads to
\begin{align}\label{constr_eng_mom_cons}
&\partial_\mu B^{\mu\nu} + m\,\partial^\nu m \int dP f \nonumber\\
+&\int dP  \,p^\nu \left[ (p\cdot \partial)f + m\, (\partial^\rho m) \, \partial_\rho^{(p)}\, f \right] = 0.
\end{align}

The Boltzmann equation for temperature-dependent particle masses can 
be written as \cite{Florkowski:1995ei, Romatschke:2011qp}
\begin{equation}\label{BE}
 (p\cdot \partial)f + m\, (\partial^\rho m) \, \partial_\rho^{(p)}\, f  = {\cal C}[f],
\end{equation}
where ${\cal C}[f]$ is the collision kernel. Substituting the above 
equation in Eq.~(\ref{constr_eng_mom_cons}), we obtain
\begin{equation}
 \partial_\mu B^{\mu\nu} + m\, \partial^\nu m \int dP f + \int dP \, p^\nu {\cal C}[f] = 0.
\end{equation}
Note that the last term in the above equation corresponds to the 
first moment of the collision kernel, which may not vanish for a 
system of quasiparticles with temperature-dependent masses\footnote 
{Quasiparticles may not be the only carriers of energy and momentum.}.

Using the equilibrium condition, Eq.~(\ref{equilibrium_bag}), we 
write the four-momentum conservation requirement for non-equilibrium 
quantities
\begin{equation}\label{eng_mom_cond}
 \partial_\mu \deltaB^{\mu\nu} + m\,\partial^\nu m \int dP \,\deltaf + \int dP \, p^\nu {\cal C}[f] = 0,
\end{equation}
where $\deltaf \equiv f-f_\eq$.  In the present calculation, we 
assume the collision kernel to be given by the relaxation-time 
approximation \cite{Anderson:1974},
\begin{equation}\label{RTA}
 {\cal C}[f] = -\frac{(u\cdot p)}{\taueq}\deltaf,
\end{equation}
where $\tau_R$ is the relaxation time. 

In order to have a well defined kinetic framework, it is necessary 
to define the fluid four-velocity $u^\mu$ and effective temperature 
$T$. Since in the present work we do not consider conserved charges, 
the natural choice for the fluid four-velocity is the Landau frame 
definition~\cite{Anderson:1974}
\begin{equation}
u_\mu T^{\mu\nu}= \ped u^\nu.
\end{equation}
It is important to note that in the present quasiparticle framework, 
the collision kernel defined in the relaxation-time approximation, 
Eq.~(\ref{RTA}), does not fix the local rest frame. Indeed, even in 
equilibrium, the mean field contribution to the energy and momentum 
$B_0$ can be significant. The quasiparticle excitations can exchange 
four-momentum with the mean fields and therefore they are not 
required to satisfy vanishing first moment of the collision 
kernel. Sources of four momentum from the kinetic contribution of 
$T^{\mu\nu}$ are acceptable as long as they are compensated with an 
opposite source on the dynamic $B^{\mu\nu}$ component.

Temperature of the system is defined through the matching condition
\begin{equation}\label{LM}
 \ped = \pedeq \quad \Rightarrow \quad \int dP (p\cdot u)^2 \deltaf  + u \cdot \deltaB  \cdot u = 0,
\end{equation}
which is equivalent to fixing the effective local temperature $T$ in 
order to reproduce the energy density of the system using the 
equilibrium relations. As a consequence, the first-moment of the 
collision kernel in the relaxation-time approximation can be written 
as
\begin{equation}
 -\frac{1}{\taueq} \int dP \,(p\cdot u) p^\nu \deltaf = \frac{1}{\taueq} \, u_\mu \deltaB^{\mu\nu}.
\end{equation}
Therefore, the condition to satisfy energy and momentum 
conservation, Eq.~(\ref{eng_mom_cond}), becomes
\begin{equation}\label{conditions_pre_ansatz}
 \partial_\mu \deltaB^{\mu\nu} + m\, \partial^\nu m \int dP \,\deltaf  + \frac{1}{\taueq} u_\mu \deltaB^{\mu\nu} = 0.
\end{equation}
After some straightforward algebra it is possible to show that for 
$\deltaB^{\mu\nu}=0$, the above equation reduces to
\begin{equation}
-3\, m \, \left( \partial^\nu m \right) \, \Pi = 0.
\end{equation}
Since we are interested in the case of temperature-dependent 
particle masses where bulk viscosity is non-vanishing, we conclude 
that, in general, $\deltaB^{\mu\nu}\neq 0$ in order to keep the 
energy and momentum conserved.

We note that in Ref.~\cite{Alqahtani:2015qja} (see also Ref.~\cite{Alqahtani:2016rth}), the Authors also 
consider the out-of-equilibrium contribution, $\deltaB^{\mu\nu}\neq 
0$, in order to obtain a set of equations for anisotropic 
hydrodynamics within a quasiparticle kinetic framework. They assume 
$\deltaB^{\mu\nu}$ to be proportional to the metric tensor, similar 
to the equilibrium case, for which the first moment of the 
collision kernel vanishes. While this is a convenient choice for 
anisotropic hydrodynamics at the leading order, where one can make 
certain approximations on the shape of the distribution function, in 
the present paper we prefer to have an extended algebra in order to 
fix $\deltaB^{\mu\nu}$ by using only the constraints obtained from 
energy-momentum conservation. Moreover, in  Refs.~\cite 
{Alqahtani:2015qja}, the definition of effective temperature out of 
equilibrium fixes the kinetic contribution to the proper energy 
density (the quasiparticle contribution, opposed to the bag 
contribution, which can be interpreted as mean fields). Since this 
kind of quasiparticle do not correspond to any known excitation, we 
prefer not to use them to define the effective temperature. We 
therefore consider a standard Landau matching in which the effective 
temperature is defined using the full energy density.

\subsection{Ansatz for $\deltaB^{\mu\nu}$ and its evolution equations}
\label{ssec2:non}

In order to specify the form of $\deltaB^{\mu\nu}$, we first note 
that, in general, the symmetry of $T^{\mu\nu}$ restricts 
$\deltaB^{\mu\nu}$ to have ten independent components. However, the 
conservation of energy and momentum leads to only four constraints. 
Therefore one has to reduce the number of independent degrees of 
freedom in $\deltaB^{\mu\nu}$ to four. In the present work, we make an \emph{ansatz} for $\deltaB^{\mu\nu}$ of the form,
\begin{equation}\label{ansatz}
 \deltaB^{\mu\nu} = b_0 \,  g^{\mu\nu} + u^\mu b^\nu + b^\mu u ^\nu,
\end{equation}
where $b^\mu$ is orthogonal to the fluid four-velocity, \textit{i.e.} \mbox{
$u\cdot b=0$}. Therefore, the matching condition, Eq.~(\ref{LM}), 
leads to
\begin{equation}
 \int dP (p\cdot u)^2 \deltaf = -b_0,
\end{equation}
while the definition of bulk viscous pressure and shear-stress 
tensor in Eqs.~(\ref{bulk_def}) and (\ref{shear_def}) reduces to
\begin{align}
 \Pi &= -\frac{1}{3}\Delta_{\alpha\beta} \int dP p^\alpha p^\beta \deltaf - b_0, \label{bulk_def_b}\\
 \pi^{\mu\nu} &= \Delta^{\mu\nu}_{\alpha\beta} \int dP p^\alpha p^\beta \deltaf. \label{shear_def_b}
\end{align}
It is interesting to observe that in the above equations, compared 
to its usual definition for a system of particles with constant 
masses, the definition of bulk viscous pressure receives a 
correction from $\deltaB^{\mu\nu}$. On the other hand, the 
definition of shear-stress tensor remains unchanged.
 
For non-vanishing masses, Eq.~(\ref{bulk_def_b}) leads to
\begin{equation}\label{Ideltaf}
 \int dP\, \deltaf = -\frac{1}{m^2}\left(\VP 3 \Pi + 4 b_0 \right).
\end{equation}
Using the above relation along with Eq.~(\ref{ansatz}) in the 
four-momentum conservation requirement for non-equilibrium 
quantities, Eq.~(\ref{conditions_pre_ansatz}), we get
\begin{align}
 \partial^\nu b_0 &+ \theta \, b^\nu +{\dot b}^\nu +(\partial\cdot b)u^\nu + (b\cdot\partial)u^\nu \nonumber\\
 &- \frac{\partial^\nu m}{m}\left(\VP 3 \Pi + 4 b_0 \right) +\frac{1}{\taueq} \left(\VP b_0 \, u^\nu + b^\nu \right) = 0,
\end{align}
where we have defined $\dot A\equiv u \cdot \partial A$ and $\theta 
\equiv\partial\cdot u$. The projection along and orthogonal to 
$u^\mu$ leads to a set of relaxation-type equations for the 
components of $\deltaB^{\mu\nu}$
\begin{align}
 {\dot b_0} +\frac{b_0}{\taueq}  =\; & \frac{{\dot m}}{m} \left(\VP 3 \Pi + 4 b_0 \right) 
 + b\cdot {\dot u} - \partial\cdot b, \label{b0_m} \\ 
 {\dot b}^{\langle\mu\rangle} +\frac{b^\mu}{\taueq} =\; & \frac{{\nabla^\mu m}}{m} 
 \left(\VP 3 \Pi + 4 b_0 \right)-\nabla^\mu b_0 -\theta \, b^\mu \nonumber \\
 &- (b\cdot \partial) u^\mu, \label{bnu_m}
\end{align}
where $\nabla^\mu\equiv\Delta^{\mu\nu}\partial_\nu$ and ${\dot 
b}^{\langle\mu\rangle}=\Delta^\mu_\nu {\dot b}^\nu$. It is 
interesting to note that in the above equations, the relaxation time 
for $b_0$ and $b^\mu$ is the same as the Boltzmann 
relaxation time.

In order to replace the $\dot m$ and $\nabla^\mu m$ by $\dot T$ and 
$\nabla^\mu T$ in Eqs.~(\ref{b0_m}) and (\ref{bnu_m}), we decompose 
the energy-momentum tensor in the hydrodynamic degrees of freedom,
\begin{equation}\label{Tmunu_hyd}
 T^{\mu\nu} = \ped u^\mu u^\nu - (\pres + \Pi)\Delta^{\mu\nu} + \pi^{\mu\nu},
\end{equation}
where ${\cal P} = {\cal P}_0(T)$ is the hydrostatic pressure (the equilibrium contribution), which depends only on the effective temperature and it is not an additional degree of freedom.
The projection of the energy-momentum conservation equation, 
$\partial_\mu T^{\mu\nu}=0$, along and orthogonal to $u^\mu$ leads to
\begin{align}
 {\dot \ped} =& - \left( \ped + \pres \right) \theta  - \Pi \, \theta 
 + \pi : \sigma, \label{energy_conservation_exact} \\ 
 \nabla^\mu \pres =& \left( \ped +\pres \right) {\dot u}^\mu 
 + \Pi \, {\dot u}^\mu -\nabla^\mu \Pi  
 + \Delta^\mu_\alpha \partial_\beta \pi^{\alpha\beta}. \label{momentum_conservation_exact}
\end{align}
where $\sigma^{\mu\nu}\equiv\Delta^{\mu\nu}_{\alpha\beta} 
\nabla^\alpha u^\beta$ is the stress tensor. Since one can always 
express $\ped$ and $\pres$ in terms of $T$ using the matching 
condition, Eq.~(\ref{LM}), and the equation of state, we can 
further rewrite the above equations as
\begin{align}
 \frac{{\dot T}}{T} =& -  c_s^2  \left(\theta  +\frac{\Pi \, \theta 
 - \pi : \sigma}{\ped + \pres} \right) , \label{Tdot_exact} \\ 
 \frac{\nabla^\mu T}{T} =&  \left( {\dot u}^\mu + \frac{ \Pi \, {\dot u}^\mu 
 - \nabla^\mu \Pi  + \Delta^\mu_\alpha \partial_\beta \pi^{\alpha\beta}}{\ped + \pres} \right),\label{nablaT_exact}
\end{align}
where we have used the thermodynamic relation, Eq.~(\ref{th_rel}), 
and the definition of squared speed of sound, $c_s^2=d\pres/d\ped$. 

Using Eqs.~(\ref{Tdot_exact}) and (\ref{nablaT_exact}) to rewrite 
$\dot m/m=\dM\,\dot T/T$ and $\nabla^\mu m/m=\dM \nabla^\mu T/T$ and 
substituting in Eqs.~(\ref{b0_m}) and (\ref{bnu_m}), we get
\begin{align}
 {\dot b_0} +\frac{b_0}{\taueq} =\;& -\dM c_s^2\left(\theta  +\frac{ \Pi \, \theta 
 - \pi : \sigma}{\ped + \pres} \right)  \nonumber \\
 & \times \left(\VP 3 \Pi + 4 b_0 \right) + b\cdot {\dot u} - (\partial\cdot b), \label{b0_exact} \\  
 {\dot b}^{\langle\mu\rangle} +\frac{b^\mu}{\taueq} =\;& \dM
 \left( {\dot u}^\mu + \frac{ \Pi \, {\dot u}^\mu -\nabla^\mu \Pi  
 + \Delta^\mu_\alpha \partial_\beta \pi^{\alpha\beta}}{\ped + \pres} \right) \nonumber \\
 &\times \left(\VP 3 \Pi + 4 b_0 \right) - \nabla^\mu b_0 -\theta \, b^\mu - (b\cdot \partial) u^\mu, \label{bmu_exact}
\end{align}
where we introduced $\dM\equiv(T/m)(dm/dT)$. Note that, at first order in 
the gradient expansion, we get $b_0=0$ and $b^\mu=0$, which is 
consistent with the results of Refs.~\cite{Sasaki:2008fg, 
Bluhm:2010qf}. Up to second-order, we find
\begin{equation}\label{b0_II}
 b_0 = -3\, \taueq \, \kappa c_s^2   \, \Pi \, \theta, \quad b^\mu = 3\, \taueq \, 
\kappa \, \Pi \, {\dot u}^\mu. 
\end{equation}

\section{Dissipative evolution equations}

In order to derive second-order evolution equations for the 
dissipative quantities, we start with the Chapman-Enskog-like 
iterative solution of the Boltzmann equation, Eq.~(\ref{BE}). The 
first-order solution is given by \cite{Romatschke:2011qp}
\begin{equation}
\deltaf_1  =  -\frac{\taueq}{(u\cdot p)}\left[ (p\cdot \partial) f_\eq + m \, (\partial_\rho m) \, 
\partial^\rho_{(p)} f_\eq \right].
\end{equation}
In the following, for simplicity, we restrict ourselves to classical 
Maxwell-Boltzmann distribution for the equilibrium distribution 
function, Eq.~(\ref{fMB}). Using Eqs.~(\ref{Tdot_exact}) and (\ref 
{nablaT_exact}), one gets~\cite{Romatschke:2011qp}
\begin{align}
 \deltaf_1 =\,&   \frac{f_\eq \,\beta \taueq}{(u\cdot p)} \!\left[\left( \frac{1}{3} p^2 
 - m^2\kappa \, c_s^2  \right) - \left( \frac{1}{3} - c_s^2 \right)\!(u\cdot p)^2\right]\! \theta \nonumber\\
 &+ \frac{f_\eq \,\beta \taueq}{(u\cdot p)} (p \cdot \sigma \cdot p). \label{deltaf1_1}
\end{align}

We can now proceed to obtain the first-order equations for the 
dissipative quantities. Using Eq.~(\ref{deltaf1_1}) in Eqs.~(\ref
{bulk_def_b}) and (\ref{shear_def_b}), and keeping terms which are 
first-order in gradients, we get
\begin{equation}\label{navier_stokes}
 \Pi = - \beta_{\Pi} \, \taueq \, \theta, \quad \pi^{\mu\nu} = 2 \, \beta_\pi \taueq \, \sigma^{\mu\nu},
\end{equation}
where
\begin{align}
 \beta_\Pi =\;& \frac{5}{3}\, \beta \, I_{3,2} - c_s^2\left( \ped + \pres \right) 
 + \dM c_s^2 \, m^2 \beta \, I_{1,1} ~,\label{beta_Pi}\\
 \beta_\pi =\;& \beta I_{3,2}, \label{beta_pi}
\end{align}
and the shear and bulk viscosities are given by the following 
relations $\beta_\pi \taueq = \eta$ and $\beta_\Pi \taueq = \zeta$, 
respectively.

In the above equations, the integral coefficients $I_{n,q}$ are 
defined as 
\begin{equation}
 I_{n,q} \equiv \frac{(-1)^q}{(2q + 1)!!}\int dP \, (u\cdot p)^{n-2q}\left(p\cdot \Delta \cdot p\right)^q f_\eq.
\end{equation}
Note that while the form of $\beta_\pi$ in Eq.~(\ref{beta_pi}) is 
identical\footnote{In the sense that it corresponds to the simple 
substitution $\beta_\pi(m,T) \to \beta_\pi (m(T),T)$, from the fixed 
mass case to the quasiparticle one.} to that obtained for the 
constant mass case \cite{Jaiswal:2014isa}, $\beta_\Pi$ in Eq.~(\ref
{beta_Pi}) is different. Using Eq.~(\ref {navier_stokes}) in 
Eq.~(\ref{deltaf1_1}), we obtain the first-viscous correction to 
the distribution function,
\begin{align}\label{deltaf1}
 \deltaf_1 =& -\frac{f_\eq \beta}{(u\cdot p) \beta_\Pi}   \nonumber \\
 & \quad \times\Bigg[\left( \frac{1}{3}p^2 
 - m^2 \kappa \, c_s^2 \right) - \left( \frac{1}{3} - c_s^2 \right)\!(u\cdot p)^2 \Bigg] \Pi\nonumber\\
 &  
 + \frac{f_\eq \beta}{ 2(u\cdot p) \beta_\pi} (p \cdot \pi \cdot p).
\end{align}
The above expression for $\deltaf_1$ will be used to derive the 
second-order evolution equations for dissipative quantities.

For the formulation of second-order viscous hydrodynamics equations 
for $\pi^{\mu\nu}$ and $\Pi$, we adopt the method developed in Ref.~\cite{Denicol:2010xn}. We consider the co-moving derivative of 
dissipative quantities from Eqs.~(\ref {bulk_def_b}) and (\ref
{shear_def_b}),
\begin{align}
{\dot \Pi} &= -\frac{1}{3} \, (u\cdot \partial) \int dP \, (p\cdot \Delta \cdot p)\, \deltaf - {\dot b_0}, \label{bulk_exact_0}\\
{\dot \pi}^{\langle\mu\nu\rangle} &= \Delta^{\mu\nu}_{\alpha\beta}\, (u\cdot\partial)
\int dP \, p^{\langle\alpha}p^{\beta\rangle} \, \deltaf, \label{shear_exact_0} 
\end{align}
where $X^{\langle\mu\nu\rangle}\equiv\Delta_{\alpha\beta}^{\mu\nu} 
X^{\alpha\beta}$ and $\dot X^{\langle\mu\nu\rangle} 
\equiv\Delta_{\alpha\beta}^{\mu\nu}\dot X^{\alpha\beta}$. 
Rearranging the Boltzmann equation, Eq.~(\ref{BE}), in the 
relaxation-time approximation, Eq.~(\ref{RTA}), we get
\begin{eqnarray}\label{BE_rearr}
 {\dot \deltaf} = -{\dot f_\eq} - \frac{1}{(u\!\cdot\! p)}\!\left[ (p\cdot \partial) f + m\, (\partial_\rho m)\, \partial^\rho_{(p)} f \right]\! 
 - \frac{\deltaf}{\taueq}.
\end{eqnarray}
Using $\deltaf=\deltaf_1$ in the right hand side of the above 
equation, substituting in Eqs.~(\ref{bulk_exact_0}) and (\ref
{shear_exact_0}) and performing the integrations, we arrive at the 
second-order evolution equations for bulk viscous pressure and shear-stress tensor
\begin{align}
\dot{\Pi} =& -\frac{\Pi}{\tau_{\Pi}}
-\beta_{\Pi}\theta 
-\delta_{\Pi\Pi}\Pi\theta
+\lambda_{\Pi\pi}\pi : \sigma, \label{BULK}\\
\dot{\pi}^{\langle\mu\nu\rangle} =& -\frac{\pi^{\mu\nu}}{\tau_{\pi}}
+2\beta_{\pi}\sigma^{\mu\nu}
+2\pi_{\gamma}^{\langle\mu}\omega^{\nu\rangle\gamma}
-\tau_{\pi\pi}\pi_{\gamma}^{\langle\mu}\sigma^{\nu\rangle\gamma}  \nonumber \\
&-\delta_{\pi\pi}\pi^{\mu\nu}\theta 
+\lambda_{\pi\Pi}\Pi\sigma^{\mu\nu}, \label{SHEAR}
\end{align}
where $\omega^{\mu\nu}\equiv\frac{1}{2}(\nabla^{\mu}u^{\nu}- 
\nabla^{\nu}u^{\mu})$ is the vorticity tensor. 

The transport coefficients obtained here are:
\begin{align} 
 \delta_{\Pi\Pi} =&\; -\frac{5}{9} \chi - \left( 1 - \dM m^2 
 \frac{I_{1,1}}{I_{3,1}} \right) c_s^2 \nonumber\\
 & + \frac{1}{3} \frac{\beta \dM c_s^2 m^2 }{\beta_\Pi} \Bigg[
  \left( 1-3c_s^2 \right) \left( \beta I_{2,1} - I_{1,1}\right) \nonumber\\
 & -\left(1  -3\dM c_s^2 \right)
m^2 \left( \beta I_{0,1} + I_{-1,1} \right) \Bigg], \label{coeff1L}\\
  \lambda_{\Pi\pi} =&\; \frac{\beta}{3\beta_\pi} \left( 2 I_{3,2} - 7 I_{3,3}  \right) 
 -  \left( 1 - \dM m^2 
 \frac{I_{1,1}}{I_{3,1}} \right) c_s^2, \label{coeff2L}\\
  \tau_{\pi\pi} =&\; 2 -  \frac{4\beta}{\beta_\pi} \, I_{3,3}, \label{coeff3L}\\
   \delta_{\pi\pi} =&\; \frac{5}{3} - \frac{7}{3} \frac{\beta}{\beta_\pi} \, I_{3,3} 
   - \frac{\beta}{\beta_\pi}\, \dM c_s^2 m^2  \left( I_{1,2}-I_{1,1} \right), \label{coeff4L}\\
  \lambda_{\pi\Pi} =&\; -\frac{2}{3} \chi, \label{coeff5L}
\end{align}
where
\begin{align}
 \chi = &\frac{\beta}{\beta_\Pi}\Bigg[ \left( 1- 3 c_s^2 \right) 
 \left(  I_{3,2}  -I_{3,1} \right) \nonumber\\
 &-  \left(1  -3\dM c_s^2 \right)
m^2 \left(I_{1,2}- I_{1,1}\right)\Bigg]. \label{chi}
\end{align}
As expected, in the constant mass limit, $\dM\to0$, the above 
transport coefficients match exactly with those obtained in 
Ref.~\cite{Jaiswal:2014isa}. Note that functions $I_{n,q}$ used 
here are equivalent to $(-1)^q I^{(R)}_{N,q}$ from Ref.~\cite 
{Jaiswal:2014isa} with $n=N-R$. 

The integral coefficients can be obtained in terms of modified 
Bessel functions of the second kind, $K_n(z)$,
\begin{align}\label{relint1}
I_{3,3} &= \frac{gT^5z^5}{210\pi^2}\!\left[\!\frac{1}{16}(K_5\!-11K_3+58K_1)-4K_{i,1}+K_{i,3}\!\right], \\
I_{3,2} &= \frac{gT^5z^5}{30\pi^2}\left[\frac{1}{16}(K_5-7K_3+22K_1)-K_{i,1}\right], \\\label{relint3}
I_{3,1} &= \frac{gT^5z^3}{2\pi^2}K_3= T (\preseq + \pedeq)=   T^2 {\cal S}_0, \\
I_{2,1} &= \frac{gT^4 z^2}{2\pi^2}  K_2(z) = \preseq +B_0 \\
I_{2,0} &=\frac{g T^4 z^2}{2\pi^2} \left[\VP3 K_2(z) + z K_1(z) \right] = \pedeq -B_0 \\
I_{1,2} &= \frac{gT^3z^3}{30\pi^2}\left[\frac{1}{4}(K_3-9K_1)+3K_{i,1}-K_{i,3}\right], \\
I_{1,1} &= \frac{gT^3z^3}{6\pi^2}\left[\frac{1}{4}(K_3-5K_1)+K_{i,1}\right], \\
I_{0,1} &= \frac{gT^2z^2}{6\pi^2}\left[\frac{1}{2}(K_2-3K_0)+K_{i,2}\right], \\
I_{-1,1} &= \frac{gTz}{6\pi^2}\left[\VP K_1-2K_{i,1}+K_{i,3}\right],
\label{relint9}
\end{align}
where the $z$-dependence of $K_n$ is implicitly understood. Here the 
function $K_{i,m}$ is defined by the integral
\begin{equation}\label{kin}
K_{i,m}(z) = \!\int_0^\infty\! \frac{d\theta}{(\cosh\theta)^m}\,\exp(-z\cosh\theta),
\end{equation}
which has the following property
\begin{equation}\label{kinkn1}
\frac{d}{dz}K_{i,m}(z) = -K_{i,m-1}(z).
\end{equation}
This identity can also be written in integral form
\begin{equation}\label{kinkn2}
K_{i,m}(z) = K_{i,m}(0) - \!\int_0^z\! K_{i,m-1}(z') dz'.
\end{equation}
We observe that by using the series expansion of $K_{i,0}(z)=K_0(z)$, 
the above recursion relation can be employed to evaluate $K_{i,m}(z)$
up to any given order in $z$.

\section{Imposing lattice QCD equation of state}
\label{ss:QPEOS}
%
In order to solve the energy-momentum conservation equations, 
Eqs.~(\ref{energy_conservation_exact}) and (\ref 
{momentum_conservation_exact}), coupled to the relaxation-type 
equations for the dissipative quantities, Eqs.~(\ref{BULK}) and (\ref
{SHEAR}), one has to impose the equation of state of the system.
This is necessary to fix the temperature dependence of the 
thermodynamic quantities like energy density and pressure, as well as 
the transport coefficients, Eqs.~(\ref {beta_Pi})-(\ref{beta_pi}) 
and Eqs.~(\ref{coeff1L})-(\ref{chi}). For this purpose it is 
sufficient to define the temperature dependence of the quasiparticle 
mass, $m(T)$, extracted for the microscopic theory in question. In 
this case one may follow the prescription employed in Refs.~\cite
{Romatschke:2011qp, Alqahtani:2015qja}.

We consider finite-temperature lQCD equation of state at zero 
baryon chemical potential computed by the Wuppertal-Budapest 
collaboration \cite{Borsanyi:2010cj}.~To correctly match the energy 
density and pressure given in Eqs.~(\ref{eng_den_B}) and (\ref
{prs_B}) with that of lQCD at asymptotically large 
temperatures (Stefan--Boltzmann limit), we fix the degeneracy factor 
to be
\begin{equation}\label{degen_fact}
g = \frac{\pi^4}{180}\left(4 (N_c^2-1)+7 N_c N_f\right),
\end{equation}
where for the number of colors and flavors we choose $N_c =3$ and 
$N_f=3$, respectively. Using Eqs.~(\ref{eng_den_B}) and (\ref
{prs_B}) one finds that the equilibrium entropy density, ${\cal 
S}_0=({\cal E}_{0}+{\cal P}_{0})/T$, as expressed in Eq.~(\ref
{th_rel}), is independent of the bag pressure, $B_0$. Therefore it 
is the most convenient quantity which can be used to determine $m(T)$
by numerically solving
%
\begin{equation}\label{mt_eqn}
\frac{g}{2\pi^2}\left(\frac{m(T)}{T}\right)K_3\!\left(\frac{m(T)}{T}\right) = \left.\frac{{\cal S}_0(T)}{T^3}\right\vert_{\text{lQCD}} ,
\end{equation}
where right hand side of the above equation is evaluated 
using lQCD results \cite{Borsanyi:2010cj}. 


For numerical convenience, we use analytic fit to the lQCD 
results for the interaction measure (trace anomaly) \cite
{Borsanyi:2010cj}
\begin{eqnarray}\label{trace} 
 \frac{{\cal I}_\text{0}(T) }{T^4}  &=& \exp\!\Big[\!-\Big(\,\!\frac{h_1}{\hT}+\frac{h_2}{\hT^2}\Big)\Big] \\
&&\hspace{-0.5cm}\times \bigg[\frac{h_0}{1+h_3 \hT^2}+\frac{f_0\big[\tanh(f_1 \hT+f_2)+1\big]}{1+g_1 \hT+g_2 \hT^2}\bigg] ,\nonumber
\end{eqnarray}
with $\hT\equiv T/(0.2 \; \rm GeV)$. Following Refs.~\cite
{Alqahtani:2015qja, Alqahtani:2016rth}, the parameters for the fit function are chosen 
as follows: $h_0=0.1396$, $h_1=-0.18$, $h_2=0.035$, $f_0=2.76$, 
$f_1=6.79$, $f_2=-5.29$, $g_1=-0.47$, $g_2=1.04$, and $h_3=0.01$. 
Using Eq.~(\ref{trace}), the equilibrium pressure is obtained by 
performing the integral
\begin{equation}
\frac{{\cal P}_{0}(T)}{T^4}=\int_0^T \frac{dT}{T}\frac{{\cal I}_\text{0}(T)}{T^4} \, ,
\label{P_func}
\end{equation}
while the energy density is calculated from the relation 
\begin{equation}
\frac{{\cal E}_{0}(T)}{T^4} = 3 \frac{{\cal P}_{0}(T)}{T^4} + \frac{{\cal I}_{0}(T)}{T^4}.
 \label{E_func}
\end{equation}
To obtain $B_0(T)$, one may use the relation expressing thermodynamic 
consistency, Eq.~(\ref{equilibrium_bag}), which reduces to
\begin{equation}\label{equilibrium_bag_bj}
\frac{d B_0(T)}{dT} = -\frac{g  T^3 z^2}{2\pi^2}  K_1\left(z\right) \frac{dm}{dT}.
\end{equation}
The above equation can be solved numerically using the boundary 
condition $B_0=0$ at $T\simeq0$. 


\begin{figure}[t]
\begin{center}
\scalebox{.48}{\includegraphics{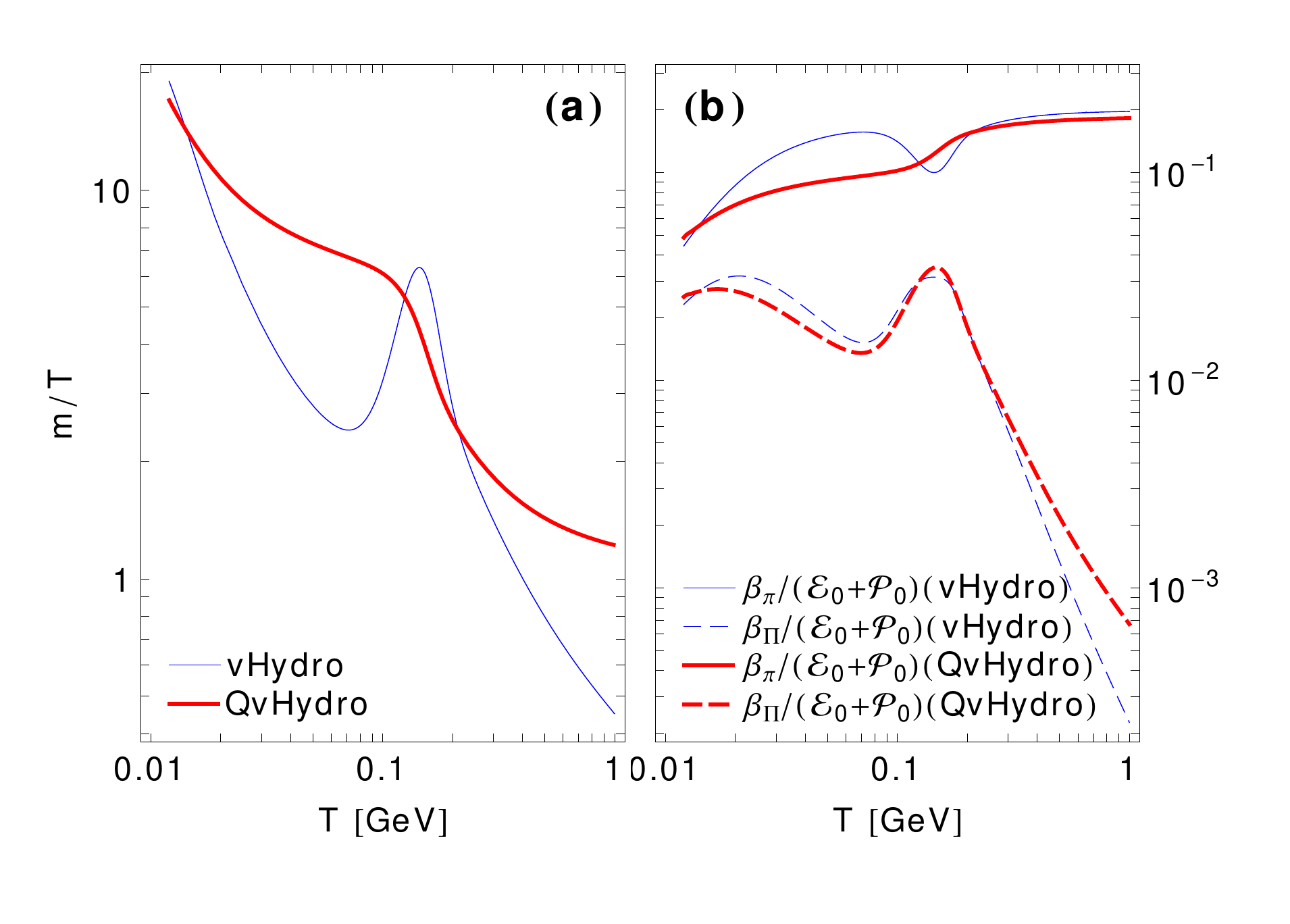}}
\end{center}
\vspace{-0.7cm}
 \caption{(Color online)  Temperature dependence of the quasiparticle mass (a) and the bulk, $\beta_\Pi$, and shear, $\beta_\pi$, first-order transport coefficients (see Eqs.~(\ref{beta_Pi}) and (\ref{beta_pi})) scaled by ${\cal E}_{0}+{\cal P}_{0}$ (b) in 
 the case of standard second-order viscous hydrodynamics with lQCD equation of state (vHydro) and the present formulation of quasiparticle second-order viscous 
 hydrodynamics (QvHydro).}
\label{Tr_coeff_visc}
\end{figure}

\begin{figure}[t]
\begin{center}
\scalebox{.48}{\includegraphics{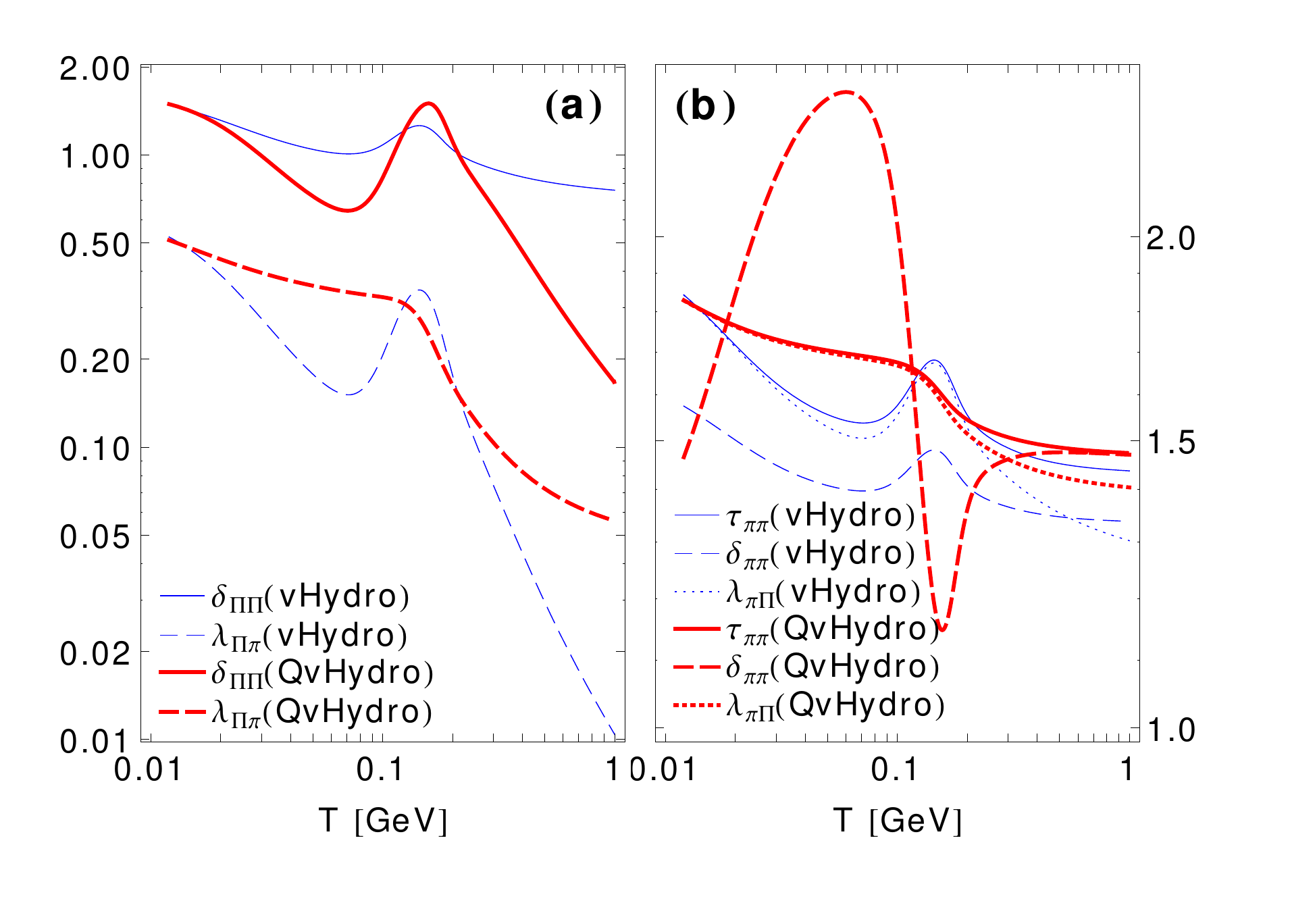}}
\end{center}
\vspace{-0.7cm}
 \caption{(Color online)  Temperature dependence of the bulk (a) and shear (b) second-order transport coefficients in 
 the case of standard second-order viscous hydrodynamics with lQCD equation of state (vHydro) and the present formulation of quasiparticle second-order viscous 
 hydrodynamics (QvHydro).}
\label{Tr_coeff}
\end{figure}

In order to compare with the usual viscous hydrodynamic results
obtained using lQCD equation of state, we use the \textit{standard}
prescription of matching speed of sound squared $c_s^2$ obtained
from kinetic theory and lQCD to extract $m(T)$. In this method, one
commonly uses a small-$m/T$ expansion for the second-order transport
coefficients \cite{Rose:2014fba, Ryu:2015vwa}. It is important to
note that this naive prescription is not only thermodynamically
inconsistent but also incompatible with the small-$m/T$ expansion;
see Fig.~\ref{Tr_coeff_visc} (a). However, using equation
$c_s^2(m(T)/T)=c_s^2(T)|_{\rm lQCD}$, one can extract $m(T)$,
exactly, via the relation \cite{Jaiswal:2014isa}
\begin{equation}\label{zex}
\frac{m(T)}{T} = \sqrt{\frac{3(\pedeq + \preseq)}{c_s^2\, \preseq}\left(\frac{1}{3}-c_s^2\right)},
\end{equation}
where, $\preseq$ and $\pedeq$ are taken from parametrization of the 
lQCD results, Eqs.~(\ref{P_func}) and (\ref{E_func}). The extracted 
$m(T)$ is then used to evaluate the transport coefficients which 
appear in the viscous evolution equations~(\ref{BULK}) and (\ref 
{SHEAR}). In the following, we refer to this prescription as the 
standard viscous hydrodynamics (vHydro).  

In Fig.~\ref{Tr_coeff_visc} (b) we 
show the temperature dependence of bulk, $\beta_\Pi$, and shear, $\beta_\pi$, first-order transport coefficients (see Eqs.~(\ref{beta_Pi}) and (\ref{beta_pi})) scaled by ${\cal E}_{0}+{\cal P}_{0}$ for 
vHydro approach (thin blue lines) and for the present framework of quasiparticle viscous 
hydrodynamics (QvHydro) (thick red lines). We observe that although the quasiparticle mass is significantly different for the case of vHydro and QvHydro formulation (see Fig.~\ref{Tr_coeff_visc} (a)), the  first-order coefficients are quite similar. In Fig.~\ref{Tr_coeff} we show the similar comparison performed for the second-order transport coefficients. In this case we see that there are substantial 
differences in the two cases for all the transport coefficients.  In 
the next Section, we study the effect of these differences in the case of
one-dimensional boost-invariant expansion of the viscous QCD medium.

%
\section{Longitudinal Bjorken flow}
%
In order to quantify the effect of the present formulation of 
quasiparticle second-order viscous hydrodynamics, we consider 
transversely homogeneous and purely-longitudinal boost-invariant, 
the so-called Bjorken, expansion \cite{Bjorken:1982qr}. The latter 
may be applicable to the early-time evolution of the viscous QCD matter in the 
very center of the heavy-ion collision.~With this symmetry it is 
convenient to use Milne coordinates, $x^\mu=(\tau,x,y,\varsigma)$, 
where $\tau\equiv\sqrt{t^2-z^2}$, $\varsigma\equiv\tanh^{-1}(z/t)$ 
and $g_{\mu\nu}=\rm{diag}(1,-1,-1,-\tau^2)$.~One may check that in 
such a case the fluid becomes static, $u^\mu=(1,0,0,0)$.

In this case the energy-momentum conservation equations, Eqs.~(\ref
{energy_conservation_exact}) and (\ref
{momentum_conservation_exact}), together with the equations for the 
dissipative quantities, Eqs.~(\ref{BULK}) and (\ref{SHEAR}), reduce to
\begin{align}
\dot{\ped} &= -\frac{1}{\tau}\left(\ped + \pres + \Pi -\pi\right) \, ,  \label{epsBj}\\
\dot\Pi + \frac{\Pi}{\tau_\Pi} &= -\frac{\beta_\Pi}{\tau} - \delta_{\Pi\Pi}\frac{\Pi}{\tau}
+\lambda_{\Pi\pi}\frac{\pi_s}{\tau} \, ,  \label{bulkBj}\\
\dot\pi_s + \frac{\pi_s}{\tau_\pi} &= \frac{4}{3}\frac{\beta_\pi}{\tau} - \left( \frac{1}{3}\tau_{\pi\pi}
+\delta_{\pi\pi}\right)\frac{\pi_s}{\tau} + \frac{2}{3}\lambda_{\pi\Pi}\frac{\Pi}{\tau} \, , \label{shearBj}
\end{align}
where $\pi_s\equiv-\tau^2\pi^{\varsigma\varsigma}$. The transport 
coefficients appearing in the above equations are given in Eqs.~(\ref{beta_Pi})-(\ref{beta_pi}) and Eqs.~(\ref{coeff1L})-(\ref{chi}), 
and the integral coefficients are given in Eqs.~(\ref{relint1})-(\ref
{relint9}). 

We note that for the results presented here we choose equal bulk and 
shear relaxation times, $\tau_\pi=\tau_\Pi= \tau_R$. Identifying the 
first-order constitutive relations, Eq.~(\ref{navier_stokes}), as 
the Navier--Stokes equations, we obtain
\begin{equation}\label{relaxt}
\tau_R=\frac{\bar{\eta}{\cal S}_0}{\beta_\pi}.
\end{equation}
In the above equation, we have defined $\bar{\eta}\equiv\eta/{\cal 
S}$ and used ${\cal S}\simeq{\cal S}_0$. One can safely make this 
approximation, because ${\cal S}={\cal S}_0+{\cal O}(\delta^2)$, 
which leads to third-order corrections in Eqs.~(\ref{bulkBj}) and 
(\ref{shearBj}) that can be ignored.

%
 
To study the evolution of the viscous QCD matter, we numerically 
solve Eqs.~(\ref{epsBj})-(\ref{shearBj}) with the transport 
coefficients given in Eqs.~(\ref{beta_Pi})-(\ref{beta_pi}) and 
Eqs.~(\ref{coeff1L})-(\ref{chi}), and supplemented with the lQCD 
equation of state, as described in Section~\ref{ss:QPEOS}. For the 
initial conditions we choose $T(\tau_i)=0.6$ GeV, $\pi_s(\tau_i)=0$ 
and $\Pi(\tau_i)=0$, where the initial proper time is $\tau_i=0.25$
fm. The initial temperature and thermalization time roughly 
correspond to those attained at the LHC. To study the effect of the 
equation of state in the entire physically interesting temperature 
range, and to compare with other results available in the literature,
we perform the evolution to extremely late times, $\tau_f=500$ fm. 
For the value of shear viscosity to entropy density ratio, we 
consider the lower bound $\eta/{\cal S}=1/(4\pi)$. We compare our 
results (QvHydro) with that obtained using quasiparticle anisotropic 
hydrodynamics (QaHydro) formulated in Ref.~\cite{Alqahtani:2015qja}, 
and the standard second-order viscous hydrodynamics (vHydro).



\begin{figure}[t]
\center
\includegraphics[height=0.25\textheight]{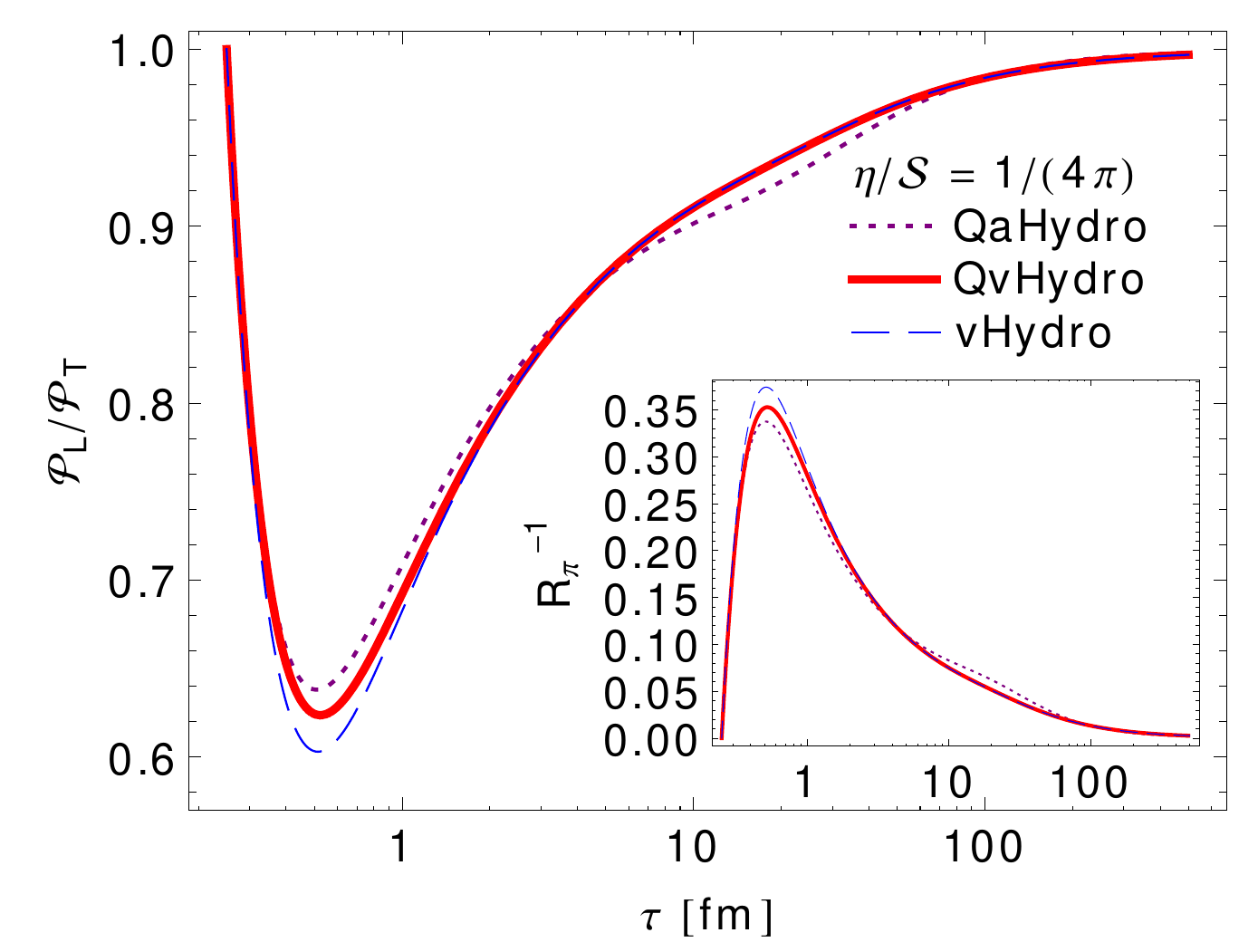}
\caption{(Color online)  The proper-time evolution of the ratio of longitudinal 
pressure to transverse pressure.  In the inset we present the respective proper-time evolution of the inverse Reynolds number $R^{-1}_\pi =\sqrt{\pi : \pi}/\preseq = \sqrt{3/2} \,\pi_s /\preseq$.} 
\label{fig:PLPT}
\end{figure} 
%
\begin{figure}[t]
\center
\includegraphics[height=0.25\textheight]{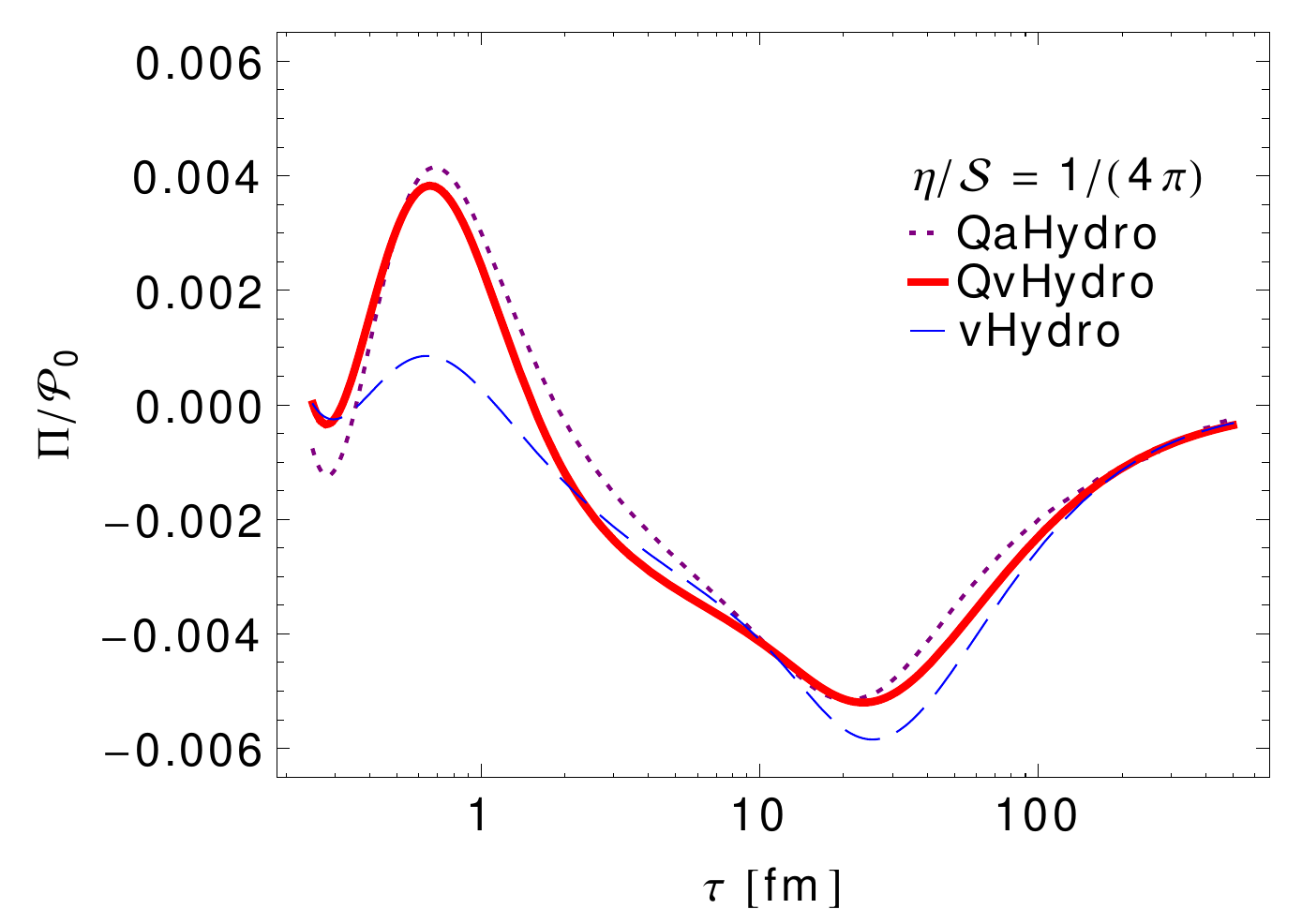}
\caption{(Color online)  The proper-time dependence of the bulk viscous pressure 
scaled by the equilibrium pressure.}
\label{fig:PiP}
\end{figure} 
%
In Fig.~\ref{fig:PLPT}, we present the proper-time evolution of the 
ratio of longitudinal pressure, ${\cal P}_L=\preseq+\Pi-\pi_s$, to 
transverse pressure, ${\cal P}_T=\preseq+\Pi+\pi_s/2$, obtained 
using QaHydro (purple dotted curve), the present formulation of 
QvHydro (red solid curve) and vHydro (blue dashed curve). We observe 
that, at early times,  the predictions concerning ${\cal P}_L/{\cal P}_T$ evolution within QvHydro and QaHydro are 
more in agreement compared to vHydro evolution. This feature is more 
evident in Fig.~\ref{fig:PiP}, where we show the proper-time 
dependence of the bulk pressure $\Pi$ scaled by the equilibrium 
pressure for the three cases. We see that, at early times, QaHydro 
and QvHydro lead to a similar increase of the scaled bulk pressure as compared to the vHydro result. We note, that the proper-time evolution of temperature resulting from these three hydrodynamic formulations is the same up to the $0.5\%$ accuracy.

\section{Summary and outlook}
%
In this paper, we have presented a first derivation of the second-order relativistic viscous 
hydrodynamics for a system of quasiparticles of a single species 
from an effective Boltzmann equation. We allowed the 
quasiparticles to have temperature-dependent masses and devised a thermodynamically-consistent framework to formulate second-order evolution equations 
for the shear and bulk viscous pressure corrections.~The formulation presented here 
is capable of accommodating an arbitrary equation of state, such as 
those obtained from lattice QCD calculations, within the framework of kinetic 
theory. It is important to maintain this consistency when one is 
using transport coefficients derived from kinetic theory for 
hydrodynamic simulations of QCD matter. Finally, we studied the effect 
of this new formulation in the case of one-dimensional purely-longitudinal 
boost-invariant expansion of viscous QCD medium formed in ultra-relativistic 
heavy-ion collisions.

Looking forward, it would be interesting to consider lQCD 
equation of state with non-zero chemical potential in the present 
calculation. This will require current conservation equation and 
evolution equation for dissipative charge current. Deriving the 
transport coefficients for particles obeying quantum statistics is 
another problem worth investigating. Moreover, since, as shown here, the thermodynamically-consistent incorporation of the realistic equation of state within the quasiparticle picture results in a significant modification of the transport coefficients, in particular bulk viscosity, it would be interesting to determine the 
impact of this change in a realistic higher-dimensional simulations. These studies may be especially interesting in the context of the recent findings concerning the importance of bulk viscosity in the evolution of matter in heavy-ion collisions \cite{Ryu:2015vwa}. We leave these 
questions for future studies.

\begin{acknowledgments}
A.J. thanks Bengt Friman and Krzysztof Redlich for helpful 
discussions.~R.R. thanks Mubarak Alqahtani, Mohammad Nopoush and Michael Strickland for discussions and 
providing results for the quasiparticle formulation of the 
anisotropic hydrodynamics, and Wojciech Florkowski for critical reading of the manuscript. The authors would like to express special 
thanks to the Mainz Institute for Theoretical Physics (MITP) for its 
hospitality and support.~A.J. was supported in part by the Frankfurt 
Institute for Advanced Studies (FIAS). R.R. was supported by Polish 
National Science Center Grant DEC-2012/07/D/ST2/02125. L.T. was 
supported by  the  U.S.  Department  of  Energy,  Office  of  
Science,  Office of Nuclear Physics under Award No. DE-SC0004286   
and  Polish National Science Center Grant DEC-2012/06/A/ST2/00390.

\end{acknowledgments}
\bibliography{references}

\end{document}